\makeatletter
\declare@file@substitution{revtex4-1.cls}{revtex4-2.cls}
\makeatother

\documentclass[twocolumn]{aastex631}
\usepackage{amsmath}
\usepackage[mathscr]{euscript}

\def\muG{{\mu\rm G}}

\received{}
\revised{}
\accepted{}

\submitjournal{The Astrophysical Journal}

\shorttitle{UHECRs from FR-I Radio Galaxies}
\shortauthors{Seo et al.}

\begin{document}

\title{Model Spectrum of UHECRs Accelerated in FR-I Radio Galaxy Jets}
\author[0000-0002-5550-8667]{Jeongbhin Seo}
\affiliation{Department of Physics, College of Natural Sciences, UNIST, Ulsan 44919, Korea}
\author[0000-0002-5455-2957]{Dongsu Ryu}
\affiliation{Department of Physics, College of Natural Sciences, UNIST, Ulsan 44919, Korea}
\author[0000-0002-4674-5687]{Hyesung Kang}
\affiliation{Department of Earth Sciences, Pusan National University, Busan 46241, Korea}
\correspondingauthor{Hyesung Kang}\email{hskang@pusan.ac.kr}
\correspondingauthor{Dongsu Ryu}\email{dsryu@unist.ac.kr}

\begin{abstract}

Nearby radio galaxies (RGs) of Fanaroff-Riley Class I (FR-I) are considered possible sites for the production of observed ultra-high-energy cosmic rays (UHECRs). Among those, some exhibit blazar-like inner jets, while others display plume-like structures. We reproduce the flow dynamics of FR-I jets using relativistic hydrodynamic simulations. Subsequently, we track the transport and energization of cosmic ray (CR) particles within the simulated jet flows using Monte Carlo simulations. The key determinant of flow dynamics is the mean Lorentz factor of the jet-spine flow, $\langle\Gamma\rangle_{\rm{spine}}$. {When $\langle\Gamma\rangle_{\rm{spine}}\gtrsim 6$, the jet spine remains almost unimpeded, but for $\langle\Gamma\rangle_{\rm{spine}}\lesssim 3$, substantial jet deceleration occurs.} CRs gain energy mainly through diffusive shock acceleration for $E\lesssim1$~EeV and shear acceleration for $E\gtrsim1$~EeV. The time-asymptotic energy spectrum of CRs escaping from the jet can be modeled by a double power law, transitioning from $\sim E^{-0.6}$ to $\sim E^{-2.6}$ around a break energy, $E_{\rm{break}}$, with an exponential cutoff at $E_{\rm{break}}\langle\Gamma\rangle_{\rm{spine}}^2$. $E_{\rm{break}}$ is limited either by the Hillas confinement condition or by particle escape from the cocoon via fast spatial diffusion. The spectral slopes primarily arise from multiple episodes of shock and relativistic shear accelerations, and the confinement-escape processes within the cocoon. The exponential cutoff is determined by non-gradual shear acceleration that boosts the energy of high-energy CRs by a factor of $\sim \langle\Gamma\rangle_{\rm{spine}}^2$. We suggest that the model spectrum derived in this work could be employed to investigate the contribution of RGs to the observed population of UHECRs.

\end{abstract}
\keywords{acceleration of particles --- cosmic rays --- galaxies: jets --- methods: numerical --- relativistic processes}

\section{Introduction}\label{s1}

Although the origin of ultra-high-energy cosmic rays (UHECRs) with $E>1~$EeV remains somewhat elusive, relativistic kpc-scale jets of radio galaxies (RGs) are considered to be the most promising source candidates \citep[see, e.g.,][for reviews]{rieger2019,matthews2020}. RG jets are commonly classified into two Fanaroff-Riley (FR) types based on their morphological characteristics \citep{fanaroff1974}: center-brightened FR-I and edge-brightened FR-II. In general, FR-I jets are observed to have lower radio luminosities and are therefore known to have lower kinetic powers compared to FR-II jets \citep[e.g.,][]{godfrey2013}. However, the morphological manifestation of RG jets looks more complicated \citep[e.g.,][]{ghisellini2001,mingo2019}; it seems to be related to the flow dynamics of jet, which depends not only on the jet power but also on the interaction with the ambient medium \citep[see, e.g.,][for a review]{hardcastle2020}.
 
With higher power, on average, individual FR-II jets are expected to produce a larger amount of UHECRs than individual FR-I jets. However, lower-power FR-I jets are more common in the universe \citep[e.g.,][]{hardcastle2019,mingo2019}. Especially in the local universe within the so-called GZK horizon of $\sim 100$~Mpc \citep{greisen1966,zatsepin1966}, virtually all RGs are the FR-I type \citep[e.g.,][]{velzen2012,rachen2019}. Among those nearby FR-I jets, several, including Virgo A (M87) and Centaurus A, exhibit blazar-like features. The inner jet of Virgo A is one-sided, hinting at the existence of bulk relativistic motions; superluminal motions found in optical and X-ray observations indicate the mean Lorentz factor of the jet-spine flow, $\langle\Gamma\rangle_{\rm{spine}}\gtrsim6$ \citep[e.g.,][]{biretta1999,snios2019}. The jet of Centaurus A is also one-sided, but the flow looks mildly relativistic with $\langle\Gamma\rangle_{\rm{spine}}\sim1.2-2$ \citep[e.g.,][]{wykes2019,snios2019b}. Meanwhile, Fornax A exhibits two inner jets embedded in the core of the host galaxy, implying recent low-power jet activity, alongside large diffuse lobes in the outer region \citep[e.g.,][]{geldzahler1984,maccagni2020}. Hence, comprehending the variety of FR-I jet structures and then understanding the processes of UHECR acceleration within jet-induced flows are crucial for modeling the observed UHECRs originating from nearby RGs.

The acceleration processes of cosmic rays (CRs) that operate in relativistic jets have been extensively investigated. They include diffusive shock acceleration (DSA) mainly at sub-relativistic shocks \citep[e.g.,][]{matthews2019}, turbulent shear acceleration (TSA) in the backflow \citep[e.g.,][]{hardcastle2010,ohira2013}, gradual shear acceleration (GSA) in the jet-spine and the backflow \citep[e.g.,][]{webb2018,rieger2019}, and non-gradual shear acceleration (nGSA) at the jet-backflow interface \citep[e.g.,][]{ostrowski1998,caprioli2015,kimura2018}. GSA and nGSA together are referred to as relativistic shear acceleration (RSA).

In \citet{seo2023} (hereafter Paper I), we studied these acceleration processes in FR-II jets by performing relativistic hydrodynamic (RHD) simulations for the flow dynamics of jets, and Monte Carlo (MC) simulations for the transport, scattering, and energy change of CR particles injected into simulated jet flows. The magnetic field is one of the key ingredients that govern physical processes in jet-induced flows; for instance, CR acceleration and synchrotron emission rely on its distribution, strength, and fluctuations. Since our numerical approach is only hydrodynamic with limited numerical resolutions, in paper I, we adopted several physically motivated prescriptions, including magnetohydrodynamic (MHD) turbulence, to model the distribution and fluctuations of magnetic fields.

Both theoretical and observational studies suggest that magnetic fields play a critical role in driving relativistic jet flow from the central engine; helical magnetic fields look to be dynamically important in inner parsec-scale jets \citep[e.g.,][]{pudritz2012}. On the contrary, magnetic fields are not dominant in the dynamics of the jet-spine flow and lobes on larger scales ranging from a few to hundreds of kpc \citep[e.g.,][]{begelman1984,laing2014}. This provides certain justification for our hydrodynamic approach to reproducing the jet-induced flow. In addition, even in the most advanced relativistic magnetohydrodynamic (RMHD) simulations, it is extremely challenging to fully resolve MHD turbulence on all the important scales \citep[see, e.g.,][]{marti2019}. Thus, it would be practical, yet scientifically justifiable, to adopt models for the distribution and fluctuations of magnetic fields and apply particle scattering laws, assuming a random walk, in MC simulations that follow the acceleration of CRs in RGs \citep[e.g.,][]{ellison1990,ostrowski1998,kimura2018}.
 
With these caveats, in Paper I, we suggested that for $E\lesssim1$~EeV, CRs gain energy mostly through DSA. But for $E\gtrsim1$~EeV, RSA becomes more important, which leads to CRs well above $10^{20}$~eV. TSA makes a relatively minor contribution over the entire energy range (see Figure \ref{f5} below). The time-asymptotic energy spectrum of CRs escaping from FR-II jets was approximated by the double-power law:
\begin{equation}
\frac{d\mathcal{N}}{dE}\propto\left(\left(\frac{E}{E_{\rm{break}}}\right)^{-s_1}+\left(\frac{E}{E_{\rm{break}}}\right)^{-s_2}\right)^{-1}, \label{spectfit}
\end{equation}
where $s_1\approx-0.5$ and $s_2\approx-2.6$. The break energy is given by the geometrical constraint known as the ``Hillas criterion''.

In this paper, we investigate the CR acceleration in FR-I jets. The flow structures in FR-I jets differ from those in FR-II jets \citep[see, e.g.,][]{perucho2014}. For example, with lower powers, in FR-I jets, $\langle\Gamma\rangle_{\rm{spine}}$ would be lower, resulting in less relativistic velocity shear in jet-induced flows. Thus, GSA and nGSA may not be as efficient as in FR-II jets. Here, we employ the same numerical approaches as in Paper I to reproduce FR-I jets and derive the energy spectrum of CRs produced at FR-I jets. We then propose a model energy spectrum for UHECRs from FR-I RGs.

The paper is organized as follows. In Section \ref{s2} and \ref{s3}, three-dimensional jet simulations and CR transport simulations are described, respectively. The results are presented in Section \ref{s4}, followed by a brief summary in Section \ref{s5}.

\begin{deluxetable*}{cccccccccccccccc}[t]
\tablecaption{Model Parameters of Simulated Jets$^a$ and Fitting Parameters for UHECR Energy Spectrum$^b$\label{t1}}
\tabletypesize{\small}
\tablecolumns{14}
\tablenum{1}
\tablewidth{0pt}
\tablehead{
\colhead{Model name$^c$}&
\colhead{$Q_j$}&
\colhead{$r_j$}&
\colhead{$\Gamma_j$}&
\colhead{$r_{\rm core}$}&
\colhead{$t_{\rm{cross}}$}&
\colhead{$t_{\rm{end}}$}&
\colhead{$s_1$}&
\colhead{$s_2$}&
\colhead{$E_{\rm{break}}$}&
\colhead{$\Gamma_{\rm{fit}}$}&
\colhead{$E_H$$^d$}&
\colhead{$E_D$$^d$}&
\colhead{$\langle\Gamma\rangle_{\rm{spine}}$$^d$}\\
\colhead{}&
\colhead{($\rm{erg~s^{-1}}$)}&
\colhead{(pc)}&
\colhead{}&
\colhead{(pc)}&
\colhead{(yrs)}&
\colhead{($t_{\rm{cross}}$)}&
\colhead{}&
\colhead{}&
\colhead{(eV)}&
\colhead{}&
\colhead{(eV)}&
\colhead{(eV)}&
\colhead{}}
\startdata
Q42-r10&3.5E+42&10&3.9&uniform&2.5E+3&50&-0.60&-2.50&2.6E+18&2.8&4.6E+18&2.3E+18&2.8\\
Q43-r10&3.5E+43&10&11.2&uniform&8.6E+2&50&-0.61&-2.65&7.5E+18&7.1&8.2E+18&7.4E+18&6.0\\
Q44-r10&3.5E+44&10&34.5&uniform&3.0E+2&75&-0.64&-2.57&1.7E+19&15.5&1.5E+19&2.3E+19&13.4\\
\hline
Q43-r32&3.5E+43&31.6&3.9&400&7.9E+3&50&-0.56&-2.51&5.7E+18&2.5&1.5E+19&6.0E+18&2.7\\
Q44-r32&3.5E+44&31.6&11.2&400&2.7E+3&50&-0.62&-2.70&2.3E+19&6.4&2.6E+19&2.2E+19&5.8\\
Q45-r32&3.5E+45&31.6&34.5&400&9.5E+2&65&-0.60&-2.35&4.3E+19&15.2&4.6E+19&6.4E+20&12.9\\
\hline
Q44-r100&3.5E+44&100&3.9&1200&2.5E+4&50&-0.56&-2.51&1.7E+19&2.5&4.6E+19&1.8E+19&2.7\\
Q45-r100&3.5E+45&100&11.2&1200&8.6E+3&50&-0.62&-2.70&7.0E+19&6.4&8.2E+19&7.0E+19&5.8\\
\hline
Q46-r1000$^e$&3.3E+46&1000&22.6&5.0E+4&2.8E+4&111&-0.45&-2.58&1.6E+20&-&1.6E+20&4.8E+20&10.3\\
\hline
\enddata
\tablenotetext{a}{Columns $2-5$: model parameters of simulated jets.}
\tablenotetext{b}{Columns $8-11$: fitting parameters for the energy spectrum of escaping CRs given in Equation~(\ref{powerexp}).}
\tablenotetext{c}{In the model name, the numbers after Q and r are the exponent of $Q_j$ and $r_j$ in units of pc, respectively.}
\tablenotetext{d}{The Hillas energy, $E_H$, the diffusion-limited maximum energy, $E_D$, and the mean Lorentz factor in the jet spine, $\langle\Gamma\rangle_{\rm{spine}}$, are the values at $t_{\rm{end}}$.} 
\tablenotetext{e}{Q46-r1000 is a FR-II jet model, presented as Q46-$\eta$5-H in Paper I.}
\vskip -0.9 cm
\end{deluxetable*}

\begin{figure*}[t]
\vskip 0 cm
\hskip 0 cm
\includegraphics[width=1\linewidth]{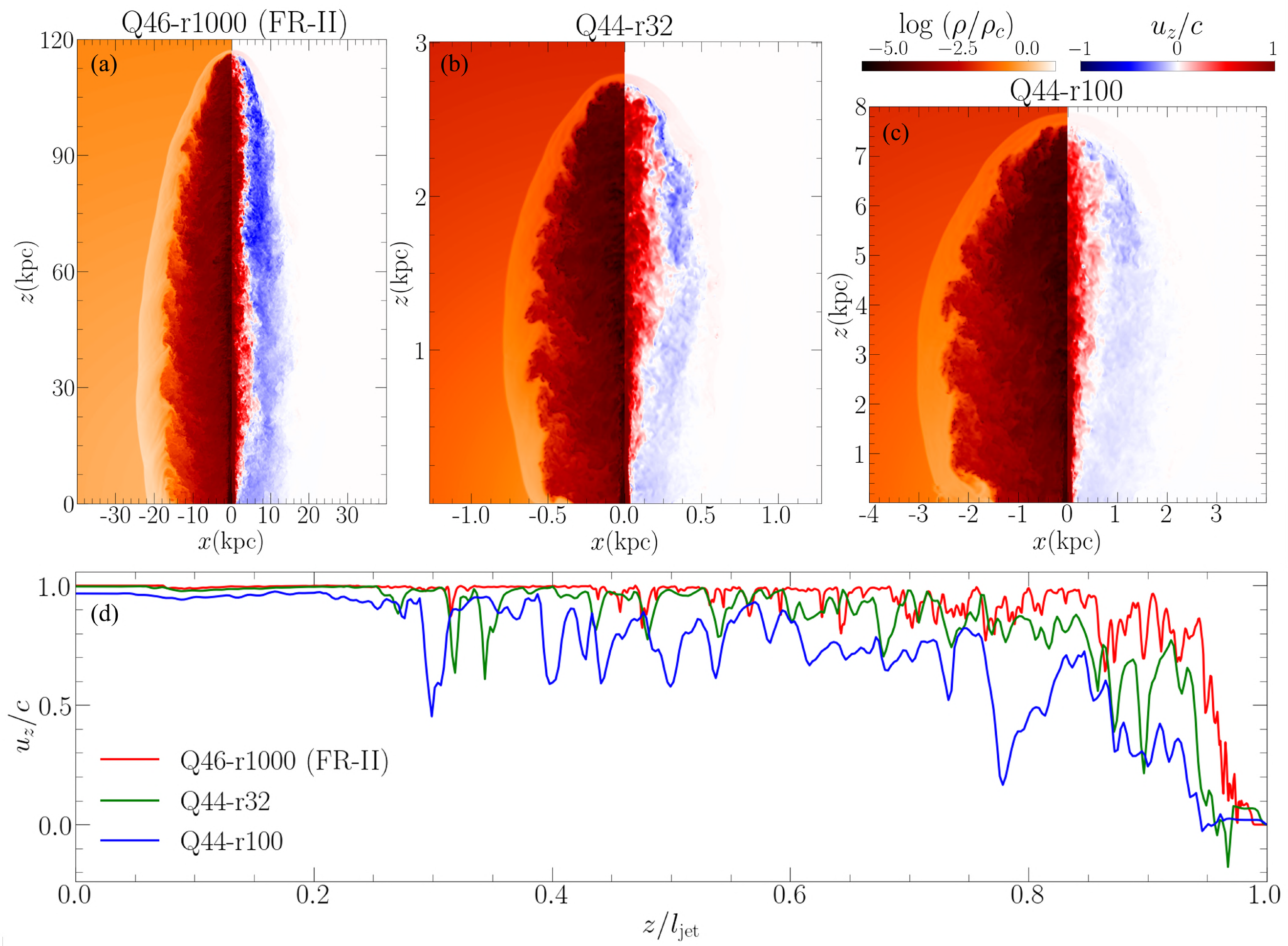}
\vskip -0.2 cm
\caption{2D slice images of the density, $\log\rho$, and the $z$-velocity along the jet direction, $u_z$, for three models, (a) Q46-r1000 (FR-II), (b) Q44-r32, and (c) Q44-r100, at $t=t_{\rm{end}}$. See Table \ref{t1} for the jet parameters. Note that the length scales of the images are different in the three panels. In Panel (d), $u_z$ along the $z$-axis is shown for the three jet models, as a function of $z/l_{\rm{jet}}$, where $l_{\rm{jet}}$ is the jet propagation length. In Q46-r1000 (red) with $\langle\Gamma\rangle_{\rm{spine}}\sim10.3$ and Q44-r32 (green) with $\langle\Gamma\rangle_{\rm{spine}}\sim5.8$, the jet-spine flows remain relatively unimpeded. On the other hand, in Q44-r100 (blue) with $\langle\Gamma\rangle_{\rm{spine}}\sim2.7$, the jet-spine flow suffers more significant deceleration. The mean Lorentz factor, $\langle\Gamma\rangle_{\rm{spine}}$, is estimated by averaging $\Gamma$ along the jet axis in the region of $z\leq2/3~l_{\rm{jet}}$.}\label{f1}
\end{figure*}

\begin{figure*}[t]
\vskip 0 cm
\hskip 0 cm
\includegraphics[width=1\linewidth]{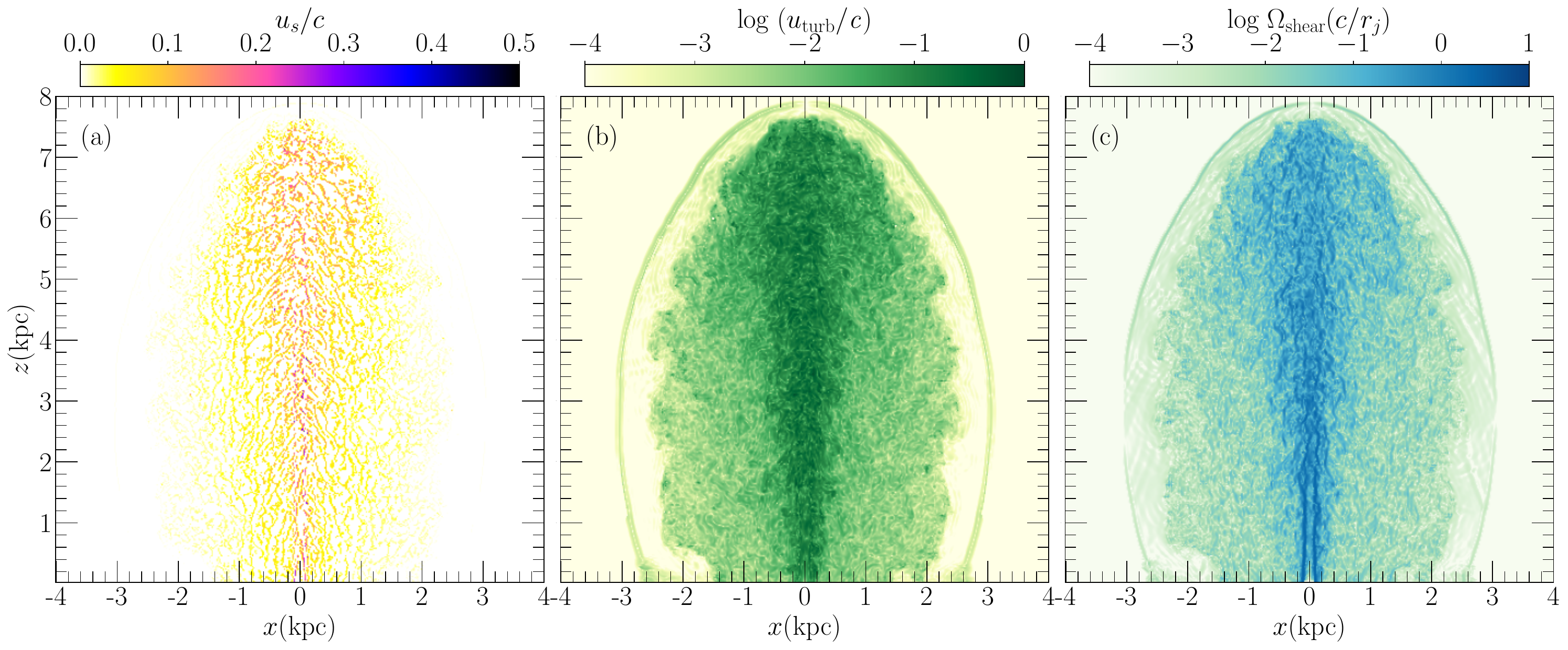}
\vskip -0.3 cm
\caption{2D slice images of the quantities that exhibit nonlinear flow dynamics in the Q44-r100 model at $t=t_{\rm{end}}$: (a) shock speed, $u_s$, (b) turbulent flow velocity, $u_{\rm turb}$, and (c) velocity shear, $\Omega_{\rm shear}$.}\label{f2}
\end{figure*}

\section{RHD Simulations for FR-I Jets}\label{s2}

\subsection{RHD Simulation Setup}\label{s2.1}

We consider relativistic jets that propagate through either the uniform or stratified background medium. The jets can be specified with the parameters of jet inflow, such as the power, $Q_j$, the density, $\rho_j$, the pressure, $P_j$, and the radius, $r_j$. Then, the bulk Lorentz factor, $\Gamma_j=1/\sqrt{1-u_j^2/c^2}$, is related to the jet power as $Q_j=\pi r^{2}_ju_{j}\left(\Gamma_j^{2}\rho_jh_j-\Gamma_j\rho_jc^2\right)$, where $u_j$, $h_j=(e_j+P_j)/\rho_j$, $e_j$, and $c$ are the jet velocity, the specific enthalpy, the sum of the internal and rest-mass energy densities, and the speed of light, respectively. {The jet flow is injected through a circular nozzle located in the bottom ($z=0$) center of the simulation box. Intending to reproduce the outer part of FR-I RGs where the jet is kinetically dominated rather than magnetically dominated, the bottom corresponds to a surface that is tens to hundreds of parsecs away from the central black hole.}

Table \ref{t1} lists the model parameters of our simulations. The first column shows the model name; the numbers after Q and r are the exponent of $Q_j$ and $r_j$ in units of pc, respectively. The models with $Q_j=3.5\times10^{42}-3.5\times10^{45}~{\rm{erg~s^{-1}}}$ and $r_j=10-100$~pc represent FR-I jets \citep[see][]{godfrey2013} that propagate up to $\sim 10$~kpc. In addition, Q46-r1000, which is the Q46-$\eta$5-H model in Paper I, is included as a FR-II jet to be compared with FR-I jets.

For FR-II jets in Paper I, the stratified, isothermal intracluster medium (ICM) was chosen as the background (see in the paper for the parameters). On the other hand, for FR-I jets in this paper, the background medium appropriate for galactic halos is employed; initially, it is assumed to be isothermal. In the models with $r_j=10$~pc, the jet propagates up to $l_{\rm{jet}}\sim1$~kpc, which is smaller than the size of typical galaxy halos. Here, $l_{\rm{jet}}$ is the jet propagation length defined by the location of the bow shock in the $z$-axis (see Figure~\ref{f1}). So the background medium is set to be uniform with the parameters relevant to the galactic environment, $\rho_0=3.0\times10^{-25}\rm{g~cm^{-3}}$ and $P_0=2.4\times10^{-10}~\rm{dyne~cm^{-2}}$, corresponding to $T=6.0\times10^6$~K \citep[e.g.,][]{perucho2014}. {In the models with $r_j= 31.6-100$~pc, the so-called King profile is adopted for the background density, $\rho_b(r)=\rho_0(1+({r}/{r_{\rm core}})^{2})^{-3\beta_K/2}$, with $r_{\rm core}=12~r_j$ and $\beta_K=0.73$ \citep{king1962}. The core radius, $r_{\rm core}$, for each model is listed in the fifth column of Table \ref{t1}.} External gravity is imposed to balance the pressure gradient due to the stratification. The background setup may be somewhat arbitrary, but the results presented in the next section are not sensitive to it. The jet-to-background density ratio is $\eta=\rho_j/\rho_0=10^{-5}$, and the pressure radio is $P_j/P_0=1$; then, $P_j/\rho_j c^2=9.3\times10^{-2}$, so initially the injected jet flow is not thermally relativistic. On the other hand, with $\Gamma_j\sim4-35$, the jet is kinetically driven.

As in Paper I, simulations are performed using the RHD code described in \citet{seo2021a}, which is based on the WENO and SSPRK schemes. To correctly produce the thermodynamics of relativistic fluids, the RC version of the equation of state is employed \citep{ryu2006}. In all the simulations, the same grid resolution of $r_j/\Delta x=5$ is used, where $\Delta x$ is the size of grid zones. The simulation results are presented in terms of the jet crossing time, $t_{\rm{cross}}={r_j}/{u_{\rm{head}}}$, where $u_{\rm{head}}\approx u_j\cdot \sqrt{\eta_r}/(\sqrt{\eta_r}+1)$ is the approximate advance speed of the jet head and $\eta_r=(\rho_{j}h_{j}\Gamma^{2}_j)/(\rho_0 h_0)$ is the relativistic density contrast \citep{marti1997}. Here, $h_0$ is the enthalpy of the background medium. Simulations run up to $t_{\rm{end}}=50-75~t_{\rm{cross}}$, and then the jet propagates typically up to $l_{\rm{jet}}\sim100~r_j$. {To break the rotational symmetry of the system, a small precession of period $10~t_{\rm{cross}}$ and angle $0.5^{\rm{o}}$ is added. We note that this is not necessarily related to the real physical precession of the inner jet from the central engine.}

\subsection{Flow structures of FR-I jets}\label{s2.2}

Figure~\ref{f1} compares the overall jet-flow dynamics and the deceleration of the jet-spine in the Q46-r1000, Q44-r32, and Q44-r100 models. Among the three, the FR-II model, Q46-r1000 with $\Gamma_j=22.6$, suffers the least deceleration, and the jet-spine flow remains relativistic almost up to the jet head. The two Q44 models have the same jet power $Q_j$ but different injection Lorentz factor $\Gamma_j$. In Q44-r32 with $\Gamma_j=11.2$, the jet-spine remains relativistic for most of its path, resulting in $\langle\Gamma\rangle_{\rm{spine}}\sim5.8$. The mean Lorentz factor, $\langle\Gamma\rangle_{\rm{spine}}$, is estimated by averaging $\Gamma$ along the jet axis in the region of $z\leq2/3~l_{\rm{jet}}$ where the jet-spine is well preserved. The values of $\langle\Gamma\rangle_{\rm{spine}}$ at $t_{\rm{end}}$ in our jet models are listed in the last column of Table \ref{t1}. In the Q44-r100 model with $\Gamma_j=3.9$, the jet flow suffers more significant deceleration, leading to $\langle\Gamma\rangle_{\rm{spine}}\sim2.7$. As a result, this jet has a broader cocoon with a larger ratio of width to length. This comparison demonstrates that, in addition to the jet power $Q_j$, the Lorentz factor $\Gamma_j$ plays a crucial role in determining the dynamics of jet flow, in particular the deceleration of the jet spine.

Figure \ref{f2} shows the two-dimensional (2D) distributions of nonlinear structures in Q44-r100, which are expected to play key rolls in CR acceleration. These include the shock speed ($u_s$), the turbulent velocity ($u_{\rm turb}$), and the velocity shear ($\Omega_{\rm shear}$). This FR-I jet initially exhibits relativistic speeds, producing strong shear flow only near the jet nozzle ($z\lesssim3$ kpc). But it soon transitions to more or less turbulent flow for $z\gtrsim3$~kpc, which subsequently generates a diffusive jet head and a broad cocoon. Shocks induced in the jet-spine and the backflow (cocoon) are subrelativistic, with $u_s\lesssim 0.2-0.3 c$.

The Q46-r1000 jet, on the other hand, induces relativistic shocks with $u_s\sim 0.4–0.5 c$ along the jet-spine flow all the way to the termination region, as shown in Figure 4 of Paper I. Compared to the Q44-r100 model, there is also stronger shear along the interface between the jet-spine and the backflow, which causes more turbulence through the Kelvin-Helmholtz instability. Hence, CRs are expected to be energized to higher energies in higher-power FR-II jets like the Q46-r1000 jet, since DSA and RSA are the dominant processes for CR acceleration in jet-induced flows (see Section \ref{s4.1} below).

\section{Monte Carlo Simulations for CR Transport}\label{s3}

The acceleration of UHECRs in FR-I jets is evaluated using Monte Carlo (MC) simulations that follow the transport and energization of particles in simulated jet flows. The interactions of particles with underlying MHD fluctuations are fundamental in determining the acceleration of CRs as well as their confinement and escape from the system. Thus, the distribution and strength of magnetic fields and the particle scattering laws are the key elements of our MC simulations. Here, we brief those, leaving the details to Paper I. A crucial measure of particle scattering and transport is the gyroradius, which depends on the local magnetic field strength $B$:
\begin{equation}
r_g \approx \frac{1.1~{\rm kpc}}{Z_i}\left(\frac{E}{ 1~{\rm EeV}}\right)\left(\frac{B}{1~\muG}\right)^{-1}, \label{rg}
\end{equation}
where $Z_i$ and $E$ are the charge and the energy of CR nuclei.

\begin{figure*}[t]
\vskip 0 cm
\hskip -0.4 cm
\includegraphics[width=1.03\linewidth]{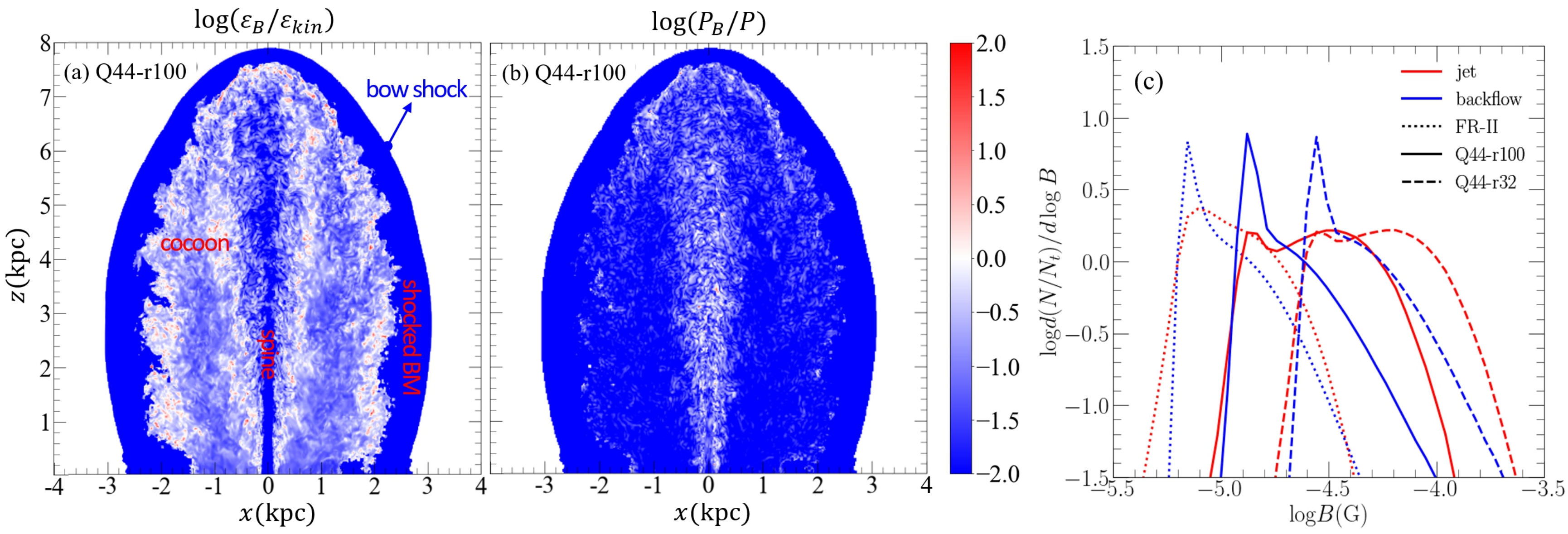}
\vskip -0.1 cm
\caption{Panel (a): Ratio of the magnetic energy density to the kinetic energy density, $\mathcal{E}_B/\mathcal{E}_{\rm kin}$, for the Q44-r100 model, where $\mathcal{E}_{\rm kin}=\Gamma_f(\Gamma_f-1)\rho c^2$ with the fluid Lorentz factor $\Gamma_f$. Panel (b): Ratio of the magnetic pressure to the gas pressure, $P_B/P$, for the same model. Here, the ``comoving'' magnetic field strength $B={\rm{max}}(B_p,B_{\rm{turb}},B_{\rm{Bell}})$ in the fluid frame is used. The two ratios are shown for the region inside the bow shock, whereas the background medium (BM) is colored white. Panel (c): Probability distribution functions of $B$ in the jet-spine flow (red) and the backflow (blue) for the FR-II (dotted lines), Q44-r100 (solid lines), and Q44-r32 (dashed lines) models. All are shown at $t_{\rm end}$.}\label{f3}
\end{figure*}

\subsection{Magnetic Field Model}\label{s3.1}

Since our RHD simulations are hydrodynamic, we adopt physically motivated model for the distribution of magnetic field. Our recipes incorporate magnetic field amplification via small-scale turbulent dynamo, $\mathcal{E}_{B,{\rm turb}}\equiv B_{\rm{turb}}^2/8\pi\approx\mathcal{E}_{\rm{turb}}$ \citep[e.g.,][]{cho2009}, and CR streaming instabilities at shocks, $P_{B,{\rm Bell}}\equiv B_{\rm{Bell}}^{2}/8\pi\approx(3/2)(u_s/c)P_{\rm{CR}}$ \citep[e.g.,][]{bell2004}. {Here, $\mathcal{E}_{\rm turb}\approx \Gamma_{\rm turb}\left(\Gamma_{\rm turb}-1\right)\rho c^2$ with $\Gamma_{\rm turb} = 1/\sqrt{1-(u_{\rm turb}/c)^2}$ is the kinetic energy density of turbulent flow;} $u_s$ is the shock speed, and $P_{\rm{CR}}$ is the CR pressure at shocks approximated as $\sim0.1\rho_1u_s^2$ with the pre-shock density $\rho_1$. In the regions of shock-free and weak turbulence, $P_{B,P}\equiv B_P^2/8\pi\approx P/\beta_p$ is adopted with the plasma beta $\beta_p\approx100$. We get $u_{\rm{turb}}$, $u_s$, $\rho_1$, and $P$ from the simulated jet flows {\citep[see][for the evaluation of $u_{\rm{turb}}$]{seo2021b}}. The highest estimate is chosen among the three model values, $B(x,y,z)={\rm{max}}(B_{\rm{turb}},B_{\rm{Bell}},B_P)$, as the local comoving magnetic field strength.

{Figure \ref{f3} (a) and (b) display the ratio of the magnetic energy density to the kinetic energy density, $\mathcal{E}_B/\mathcal{E}_{\rm kin}$, and the ratio of the magnetic pressure to the gas pressure, $P_B/P$, respectively. Here, $\mathcal{E}_{\rm kin}=\Gamma_f(\Gamma_f-1)\rho c^2$, and $\Gamma_f=1/\sqrt{1-(u/c)^2}$ is the fluid Lorentz factor calculated with the fluid speed $u$. The figures show that the magnetic field is subdominant, that is, either $\mathcal{E}_B<\mathcal{E}_{\rm kin}$ or $P_B<P$ in most regions. In our magnetic field model, $B$ is set mostly by $B_{\rm{turb}}$ in the jet spine as well as in the dynamically active parts of the cocoon, such as the interfaces between the jet-spine and the cocoon and between the cocoon and the shocked background medium. On the other hand, in relatively quiet regions of the cocoon, $B$ is fixed mostly by $B_P$.  Figure \ref{f3} (c) plots the probability distribution functions (PDFs) of the magnetic field strength in the jet-spine flow (red lines) and the cocoon (blue lines) for three models, Q46-r1000 (FR-II), Q44-r100, and Q44-r32. The peaks represent quiet regions, while the broad distributions mainly include dynamically active parts. The magnetic field is stronger if $\mathcal{E}_{\rm{turb}}$ and $P$ are larger.}

The magnetic field strength in the observer frame can be approximated as $B_{\rm obs}\approx \Gamma_f\cdot B(x,y,z)$; then typical values range $B_{\rm obs}\sim 10-30 \mu$G in the backflow and $B_{\rm obs}\sim 10-100 \mu$G in the jet-spine flow in our model. These estimates lie within the range of values from X-ray and radio observations \citep[e.g.,][]{begelman1984,kataoka2005,anderson2022}. The general features of the spatial distribution of $B_{\rm obs}$ for a FR-II jet are shown in Figure 3(b) of Paper I. These features also apply to lower-power FR-I jets.

{Our magnetic field model employs $\mathcal{E}_{\rm B,turb}/\mathcal{E}_{\rm turb}\approx1$ and $\beta_p\approx100$. While they result in $B_{\rm obs}$ consistent with observations, these values are somewhat arbitrary. Hence, as comparison models, we consider the cases of $\mathcal{E}_{\rm B,turb}/\mathcal{E}_{\rm turb}\approx1/2$ and $\beta_p\approx10$; the former gives rise to a weaker $B$ than the fiducial model, while the latter gives a stronger $B$. With these, we examine the effects of magnetic field modeling on the acceleration of UHECRs in Section \ref{s4.5}.}

\subsection{Particle Scattering Laws}\label{s3.2}

CR particles are assumed to scatter elastically off underlying fluctuating magnetic fields on all relevant scales that are frozen in the local flow. The scattering is modeled with Bohm diffusion \citep[e.g.,][]{bohm1949}, Kolmogorov-type resonant scattering \citep[e.g.,][]{stawarz2008}, or nonresonant scattering \citep[e.g.,][]{sironi2013}. In general, the scattering mean free path (MFP) can be formulated as
\begin{equation}
\lambda_f(E)=L_{0}\left(\frac{E}{E_{H,L_0}}\right)^{\delta},\label{mf}
\end{equation}
where $L_0$ is the coherence length scale of fluctuating magnetic fields. The {\it local} Hillas energy is derived from the confinement condition, $r_g\approx L_0$,
\begin{equation}
E_{H,L_0}\approx 0.9~{\rm{EeV}}\cdot Z_i\left(\frac{B}{1\mu\rm{G}}\right)\left(\frac{L_0}{1{\rm{kpc}}}\right),
\label{EHL0}
\end{equation}
where $B(x,y,z)$ is the magnetic field strength in the local fluid frame (i.e., scattering frame). We set the charge $Z_i=1$ since only proton is considered in this work. We approximate $L_0\approx r_j$, as the coherence length scale of induced turbulence would be comparable to the jet radius \citep{seo2021b}.

As the fiducial model, we adopt $\delta=1/3$ (Kolmogorov) for $E\leq E_{H,L_0}$ and $\delta=1$ (Bohm) for $E>E_{H,L_0}$, but $\delta=1$ for all energy around shocks. We also consider $\delta=1/3$ and $2$ (nonresonant) for $E>E_{H,L_0}$ as comparison cases. {In addition, as another comparison case, we consider the diffusive transport model in turbulent magnetic fields presented by \citet{harari2014}:
\begin{equation}
\begin{aligned}
\lambda_f(E)\approx\frac{L_{0}}{a_H+a_I+a_L}~~~~~~~~~~~~~~~~~~~~~~~~~~~~~~~~~~~~~~~~~~\\
\times\left[ a_H\left(\frac{E}{E_{H,L_0}}\right)^2 + a_I\left(\frac{E}{E_{H,L_0}}\right)+ a_L\left(\frac{E}{E_{H,L_0}}\right)^{1/3} \right],\label{harari}
\end{aligned}
\end{equation} 
where the coefficients are set as $a_H=4$, $a_I=0.9$, and $a_L=0.23$ for the Kolmogorov spectrum. Here, the first term $\propto E^2$ represents the nonresonant scattering at high energies in the quasirectilinear propagation regime, while the third term $\propto E^{1/3}$ corresponds to the resonant scattering at low energies in the diffusive scattering regime. The middle term $\propto E$ is introduced to describe a smooth transition from low to high energies. This MFP model is similar to our model with $\delta=2$ for $E>E_{H,L_0}$, except the smooth transition around $E_{H,L_0}$.}

After scattering, ``restricted'' random walks are applied; high-energy particles with $\lambda_f(E)\gtrsim L_{0}$ move in mostly forward directions, while particles with $\lambda_f(E)\lesssim L_{0}$ scatter almost isotropically (see Paper I for details). We also consider fully isotropic scattering as a comparison case. See Section \ref{s4.5} for the differences due to different particle scattering models.

\subsection{Injection of Galactic CRs}\label{s3.3}

\begin{figure}[t] 
\vskip -0.1 cm
\hskip -0.6cm
\includegraphics[width=1.1\linewidth]{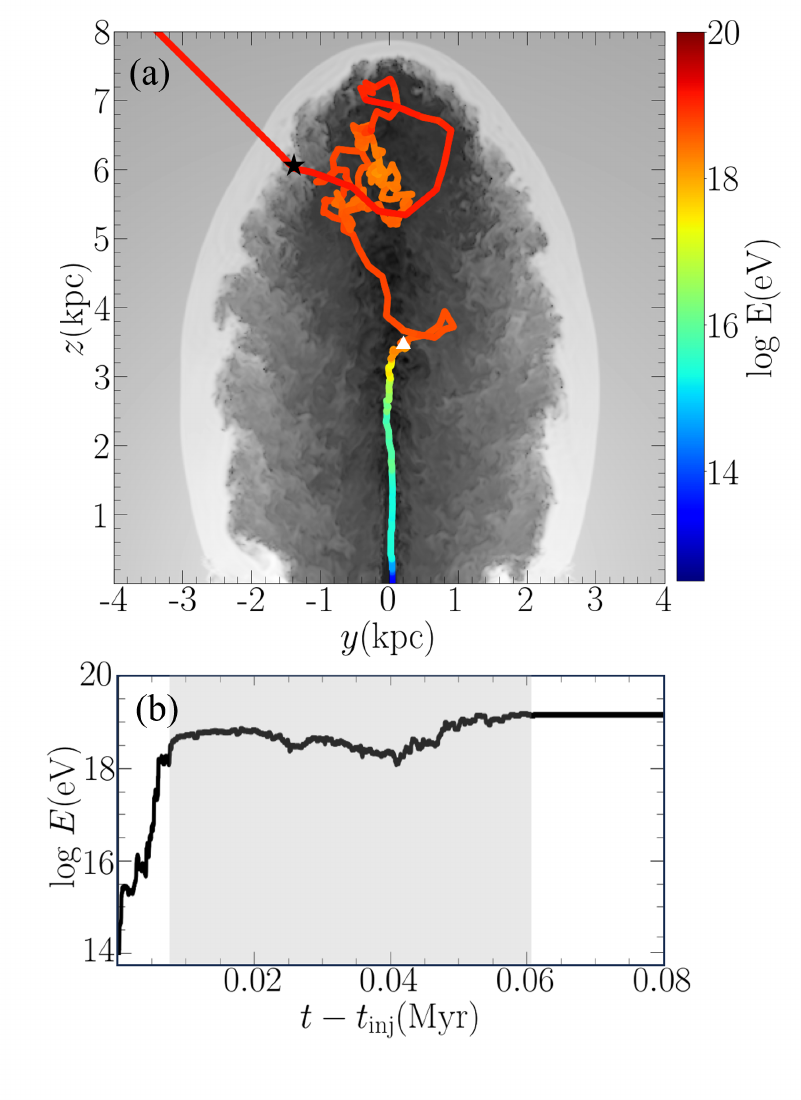}
\vskip -0.9 cm
\caption{Panel (a): Trajectory of a sample particle since the injection into the jet flow, superimposed on the distribution of $\log \rho$ in the Q44-r100 model. The trajectory is color-coded by the energy of the particle. Panel (b): Energization history of the same particle along the trajectory. The shaded region indicates the period during which the particle scatters within the diffusive head region. The starting and ending locations of that period are marked by the white triangle and black star symbols, respectively, in panel (a).}\label{f4}
\end{figure}

\begin{figure*}[t]
\vskip 0 cm
\hskip -0.4 cm
\includegraphics[width=1.03\linewidth]{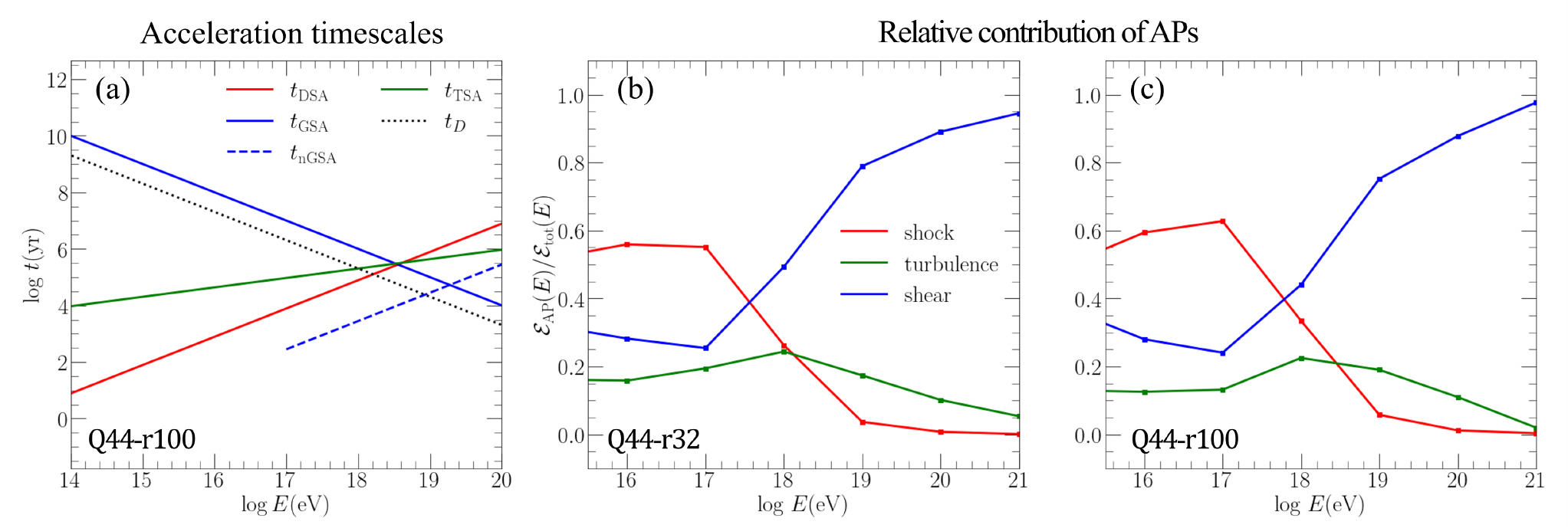}
\vskip -0.2 cm
\caption{Panel (a): Timescales of the acceleration processes (APs) that energize CRs in jet-induced flows as a function of the energy of particles for the parameters relevant to the Q44-r100 model. nGSA is operative only when $\lambda_f$ is large enough to cross the jet-backflow interface, and hence $t_{\rm{nGSA}}$ (blue dashed line) is drawn for $E>0.1$~EeV. The diffusion timescale, $t_D$, (black dotted line) is also shown for comparison. Panels (b-c): Fractions of the cumulative energy gains due to different APs as a function of the energy of particles escaping from the jet system for two models, (b) Q44-r32 and (c) Q44-r100. See the text for details.}\label{f5}
\end{figure*}

Seed CRs are injected into the jet inflow with an initial spectrum of $d\mathcal{N}_{\rm{inj}}/dE_{\rm{inj}}\propto E_{\rm{inj}}^{-2.7}$ for $E_{\rm{inj}}=10^{13}-10^{15}$~eV. They represent a population of galactic CRs accelerated within the host galaxy. {While this injection model including the slope and range is somewhat arbitrary, we found that the spectrum of escaping UHECRs doesn't depend on it at all. This is because, in our MC simulations, the particle energy increases by a factor of $\sim 10^5$; once we inject a sufficient number of galactic CRs, we obtain a smooth spectrum of escaping UHECRs up to $\sim10^{21}$eV. Hence, in effect, the injected spectrum behaves almost like a delta function in energy. Since CRs are treated as test particles, the amplitude of the resulting energy spectrum scales linearly with the number of injected CRs.}

The trajectories of CRs are followed according to the prescribed models for scattering MFP and random walks. Figure \ref{f4} illustrates the trajectory and energization history of a sample particle. Upon each scattering, a net energy change, $\Delta E$, arises as a result of the Lorentz transformation between the moving fluid frame and the simulation (laboratory) frame. In this example, the particle gains energy mainly through multiple episodes of DSA and GSA up to $\sim 1$ EeV, advecting along the jet-spine flow. Then, it undergoes diffusive scatterings in the cocoon (between the triangle and star symbols in Panel (a)) and eventually escapes from the turbulent head region. As shown in the shaded box in Panel (b), the acceleration is rather inefficient for 0.01 $\lesssim t - t_{\rm inj} \lesssim$ 0.06, during which the particle experiences diffusive scatterings within the upper cocoon ($z\gtrsim 3$~kpc).

\section{Result}\label{s4}

\subsection{Relative Importance of Acceleration Processes}\label{s4.1}

Similar to FR-II jets, various CR acceleration processes, including DSA, TSA, GSA, and nGSA, are at play within FR-I jets. We first utilize the acceleration timescales, $t_{\rm{DSA}}$, $t_{\rm{TSA}}$, $t_{\rm{GSA}}$, and $t_{\rm{nGSA}}$ (see Eqs. [11], [14], [15], and [16] in Paper I) to compare the relative efficiencies of these acceleration processes (APs). For illustrative purposes, these timescales are displayed in Figure~\ref{f5}(a) for the Q44-r100 model with the following representative parameters: the shock compression ratio, $\chi\sim4$, $u_s/c\sim0.1$, $B\sim30~\mu$G, the turbulence speed, $|u_{\rm{turb}}|/c\sim0.1$, $L_0\sim0.1$~kpc, the velocity shear, $\Omega_{\rm{shear}}=\partial u_z/\partial r\sim 0.05~c/r_j$ with the $z$-velocity along the jet direction, $u_z$, $\Gamma_z=1/\sqrt{1-u_z^2/c^2}\sim1$, and $\delta=1$ for MFP. The figure indicates that seed CRs would gain energies at first via DSA, and then RSA (GSA and nGSA) becomes increasingly significant above $E\sim 1$~EeV. The particles with $\lambda_f(E)\lesssim r_j$ stay within the shear layer spread over the jet-spine and the backflow, and gain energy mainly via GSA. In contrast, those with $\lambda_f\gtrsim r_j$ can be scattered across the jet-backflow interface and boosted to higher energies via nGSA.

The relative contributions of APs in our jet models are estimated as follows. For each scattering event, a fraction of $\Delta E$ is distributed among different APs with weight $\xi_{\rm{AP}}=t_{\rm{AP}}^{-1}/\sum_{\rm{AP}}{t_{\rm{AP}}^{-1}}$, where the summation includes DSA (shock), TSA (turbulence), and GSA (shear); nGSA is not considered separately since only a small fraction of high-energy particles undergo it. For all particles escaping from the system up to $t_{\rm{end}}$, whose final energy lies in the logarithmic bin of $[\log E,\log E+d\log E]$, $\xi_{\rm{AP}}\Delta E$ is summed to make the cumulative energy gain, $\mathscr{E}_{\rm{AP}}(E)$, and $\mathscr{E}_{\rm{tot}}= \mathscr{E}_{\rm{shock}}+\mathscr{E}_{\rm{turb}}+\mathscr{E}_{\rm{shear}}$. Figures~\ref{f5}(b)-(c) show $\mathscr{E}_{\rm{AP}}(E)/\mathscr{E}_{\rm{tot}}$ for the two Q44 models presented in Figure~\ref{f1}. The figures confirm that whereas DSA is dominant for $E\lesssim1$~EeV, shear acceleration becomes important beyond $E\gtrsim 1$~EeV. TSA plays a secondary role.

\subsection{Energy spectrum of CRs from FR-I Jets}\label{s4.2}

\begin{figure}[t]
\vskip 0 cm
\hskip -0.3 cm
\includegraphics[width=1.03\linewidth]{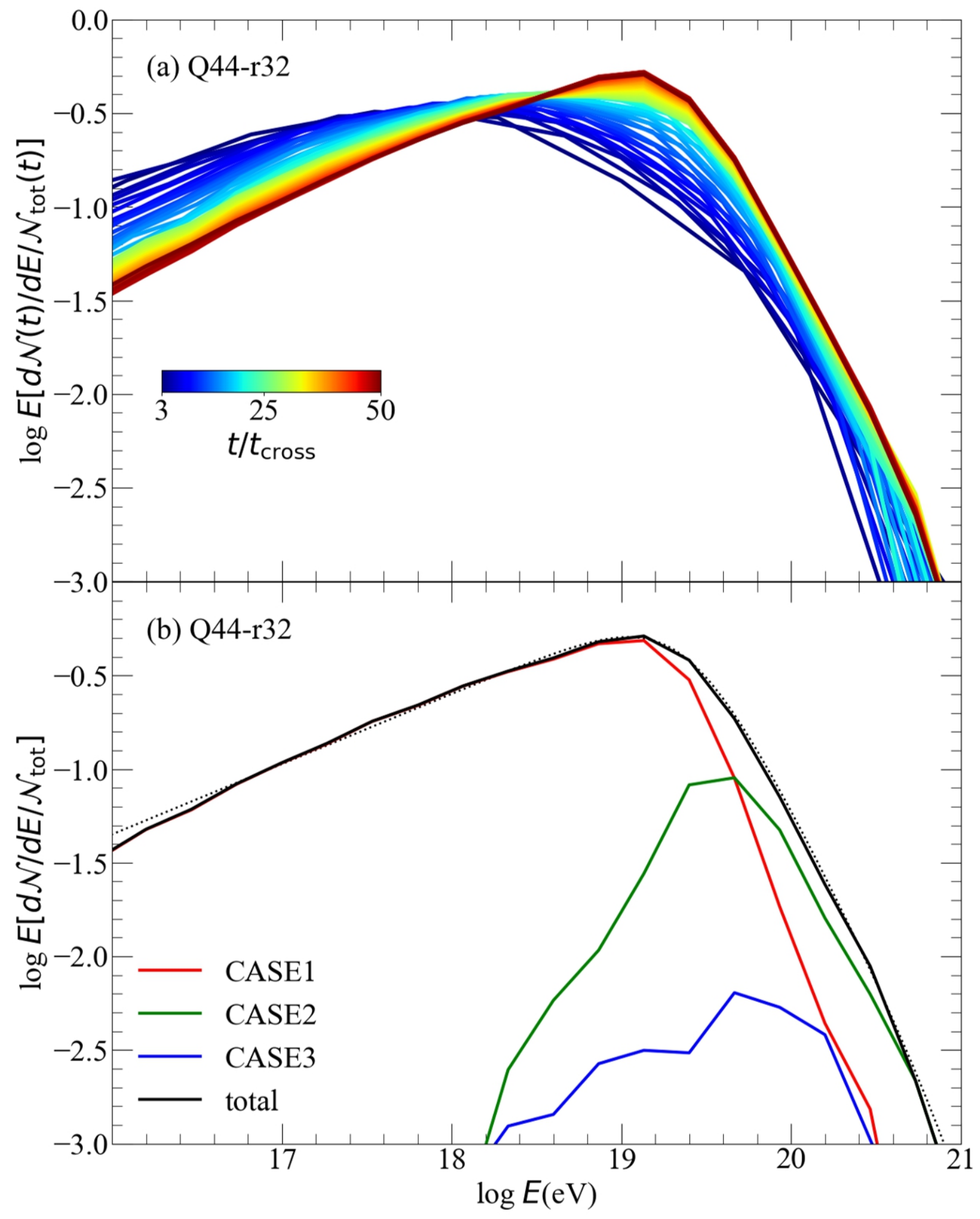}
\vskip -0.2 cm
\caption{Panel (a): Time-integrated energy spectrum, $E[d\mathcal{N}/dE/\mathcal{N}_{\rm{tot}}]$, of all particles escaping from the system up to a given time, $t$, in the Q44-r32 model. Here, $\mathcal{N}_{\rm tot} = \int (d\mathcal{N}/dE) dE$. The lines are color-coded from blue to brown, based on the time, $t/t_{\rm cross}$. Panel (b): Time-asymptotic, cumulative energy spectra, drawn separately for particles classified as CASE 1, CASE 2, and CASE 3, in Q44-r32. See the main text for the description of CASEs. The back solid line shows the total spectrum, and the black dotted line is the fitting to it.}\label{f6}
\end{figure}

The energy spectrum of the CRs that are accelerated and escape from the jet system is typically time-dependent, transitioning from an ``age-limited'' regime (bluish lines) to a ``size-limited'' regime (reddish lines), as shown in Figure \ref{f6}(a). During the early stage, all particles are accelerated mainly by encountering multiple shocks and gradual velocity shear in the jet-spine flow (see Figures \ref{f4}), and the spectrum extends to higher energies over time as the maximum energy of CRs increases with the jet's age. In the late stage, particles are confined in the jet system and continuously energized until they escape. The shape of the time-integrated, cumulative energy spectrum of all particles escaping from the jet converges to its time-asymptotic form.

{Figures \ref{f1} and \ref{f2} show that the jet spine typically disperses over a width of $\sim {\rm several} ~r_j$, and the corrugated interface between the upward-moving jet flow and the downward-moving backflow spreads over of order $\sim r_j$. Only UHECRs energized to $E\gtrsim1$~EeV have large enough MFPs to cross the shear layer between the jet-spine and the backflow, as mentioned above. Hence, as CRs are initially energized mainly via DSA, GSA first becomes important, and then UHECRs with $E\gtrsim1$~EeV cross and recross the shear interface and are energized via nGSA; each cycle of cross-recross increases the energy by a factor of $(\Gamma_\Delta^2-1)$ \citep[e.g.,][]{rieger2019,caprioli2015,mbarek2019}.} Here, $\Gamma_\Delta$ is the Lorentz factor of the velocity jump across the shear layer.

In Paper I, we categorized UHECRs produced at RGs into three types, depending on whether they experience an additional boost via nGSA before escape and how they escape from the jet structure (see Figure~9 in Paper I). In CASE 1, particles are energized mainly via DSA and RSA, and then escape from the cocoon without undergoing an additional boost via nGSA. {In CASE 2, after gaining energies via DSA and RSA, particles undergo an additional boost and then escape from the cocoon.} CASE 3 is the same as CASE 2, except that UHECRs escape directly from the jet-spine into the background medium. In Figure~\ref{f6}(b), the time-asymptotic, cumulative energy spectra, $\mathcal{N}(E)$, of the particles categorized into different CASEs are separately plotted for the Q44-r32 model at $t_{\rm{end}}$. {Most particles belong to the CASE 1 category, while only a small fraction becomes CASE 2 and 3 by undergoing an additional boost to higher energies via nGSA before escape.}

\begin{figure*}[t]
\vskip 0 cm
\hskip -0.4 cm
\includegraphics[width=1.03\linewidth]{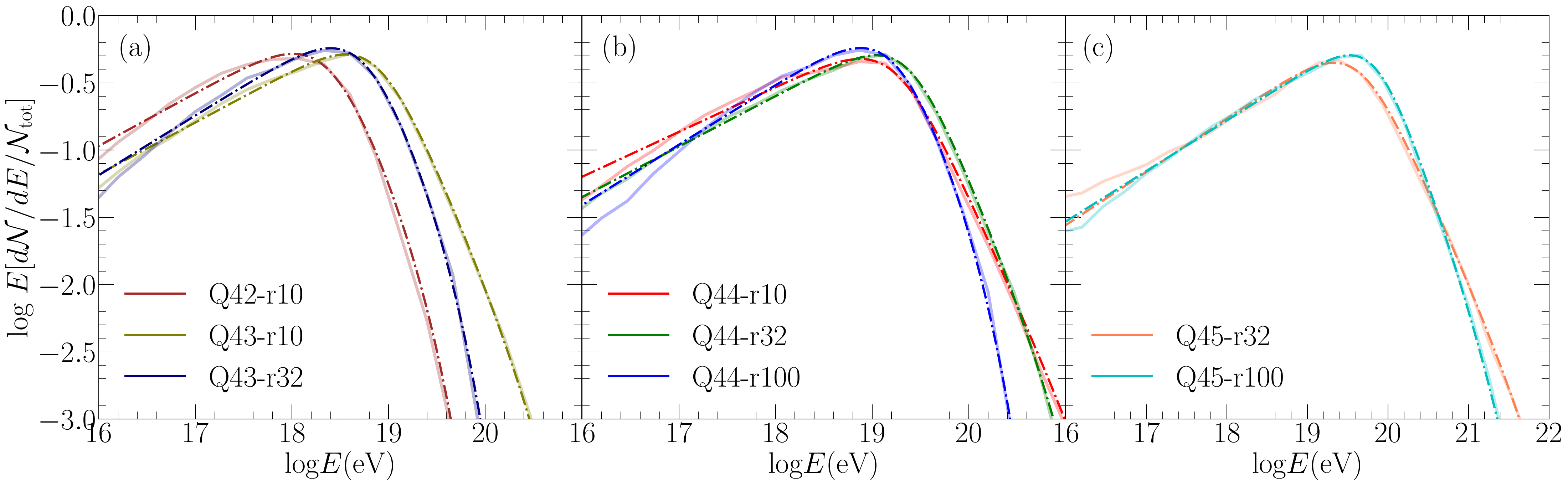}
\vskip -0.2 cm
\caption{Time-asymptotic, cumulative energy spectra of all particles escaping from the jet system up to $t_{\rm{end}}$, $E[d\mathcal{N}/dE/\mathcal{N}_{\rm{tot}}]$, (a) for the Q42 and Q43 models, (b) for the Q44 models, and (c) for the Q45 models. Each spectrum is normalized with $\mathcal{N}_{\rm{tot}}=\int(d\mathcal{N}/dE)dE$. The results of MC simulations are shown with color-coded solid lines, while the fittings to Equation~(\ref{powerexp}) are overlaid with dot-dashed lines. See Table \ref{t1} for the jet and fitting parameters. Note that the abscissa in Panel (c) stretches to a higher energy than in other panels.}\label{f7}
\end{figure*}

Figure~\ref{f7} shows $\mathcal{N}(E)$ of all escaping particles for different FR-I jet models at $t_{\rm{end}}$ with color-coded solid lines. The spectra shift to higher energies for models with higher $Q_j$, while the models with the same $Q_j$ but larger $\Gamma_j$ have spectra that extend to higher energies. In contrast to the double-power law in Equation~(\ref{spectfit}) for FR-II jets, the fitting of the CR energy spectra for FR-I jets requires an exponential cutoff:
\begin{equation}
\begin{aligned}
\frac{d\mathcal{N}}{dE}\propto\left(\left(\frac{E}{E_{\rm{break}}}\right)^{-s_1}+\left(\frac{E}{E_{\rm{break}}}\right)^{-s_2}\right)^{-1}\\
\times\exp\left(-\frac{E}{E_{\rm{break}}\Gamma_{\rm{fit}}^2}\right).~~~~~~~~~~~~~~~~\label{powerexp}
\end{aligned}
\end{equation}
{
Table \ref{t1} lists the four fitting parameters, $s_1$, $s_2$, $E_{\rm{break}}$, and $\Gamma_{\rm{fit}}$, with which Equation (\ref{powerexp}) could approximate the energy spectra of escaping CRs obtained from our MC simulations; the fittings are shown with dot-dashed lines in Figure~\ref{f7}.}

\subsection{Spectral Slopes and Exponential Cutoff}\label{s4.3}

The slopes, $s_1$ and $s_2$, in Table \ref{t1} exhibit relatively weak dependence on jet models; they are, on average, $s_1\approx-0.6$ below the break energy, $E_{\rm break}$, and $s_2\approx-2.6$ above the break energy. These values for FR-I jets are almost identical to those for FR-II jets presented in Paper I, since the two types of RG jets share similar flow dynamics and hence the same acceleration processes operate.

The power law for $E<E_{\rm break}$, $\sim E^{-0.6}$, is fixed mostly by CASE 1 particles, as shown in Figure \ref{f6}(b); we interpret it as a consequence of multiple episodes of DSA and RSA in the jet-spine flow. Multiple shock passages are known to lead to a $E^{-1}$ energy spectrum for a large number of shocks, independent of the shock compression ratio \citep[e.g.,][]{melose1993,kang2021}. {Moreover, MC simulation studies by \citet{ostrowski1998} and \citet{kimura2018} showed that the energy spectrum of particles accelerated via nGSA scales as $E^{-1}-E^0$. They used a simpler, cylinder-shaped jet-cocoon system with discrete velocity shear as the background flow for their MC simulations, instead of time-evolving, complex jet-induced flows with lots of shocks and turbulence, as seen in Figure \ref{f2}. The slope $-0.6$ is close to that of multiple DSA and also in the range of the values of \citet{ostrowski1998} and \citet{kimura2018}. Our energy spectra are harder than the canonical single DSA spectrum of $E^{-2}$.}
 
The behavior of the spectra for $E>E_{\rm break}$, on the other hand, is the combined result of particles of all CASEs (Figure \ref{f6}(b)). {In the conventional Hillas picture in which CRs are confined by an isotropic system with mean magnetic field strength $\langle B \rangle$ and size $L$, the spectrum of escaping particles would decrease exponentially above the Hillas energy (Equation (\ref{EHL0}) with $\langle B \rangle$ and $L$).} In the case of jet-induced flows under consideration, the escape scenario is modified by the following two factors: particles are confined in the {\it elongated} cocoon, and some UHECRs receive the $\langle\Gamma\rangle_{\rm{spine}}^2$ boost via nGSA and reach above the Hillas energy before escaping from the jet. {This picture is consistent with the so-called espresso acceleration model proposed by \citet{caprioli2015} and \citet{mbarek2019}.} As we mentioned in Paper I, the particle escape from an idealized, infinite cylindrical volume leads to an energy spectrum that scales as $E^{-2}$. Somewhat steeper spectra with $\sim E^{-2.6}$ are obtained in our simulations since our jets have finite-sized, barrel-shaped cocoons.

At the highest energy end, an exponential cutoff would appear at $E_{\rm{break}}\langle\Gamma\rangle_{\rm{spine}}^2$, as a consequence of the nGSA boost. Indeed, the values of $\Gamma_{\rm{fit}}$ in the exponential cutoff of Equation (\ref{powerexp}) are close to those of $\langle\Gamma\rangle_{\rm{spine}}$, as can be seen in Table \ref{t1}. Hence, it is $\langle\Gamma\rangle_{\rm{spine}}$ that primarily governs the highest energy end of the spectrum. When $\langle\Gamma\rangle_{\rm{spine}}\gtrsim10$, as in the case of FR-II jets, the cutoff energy is sufficiently high, so the highest energy part of the energy spectrum would resemble a power law. In contrast, with $\langle\Gamma\rangle_{\rm{spine}}\lesssim6$ in our simulations, the energy spectrum for FR-I jets needs an exponential cutoff.

\subsection{Break Energy of Power-Law}\label{s4.4}

The break energy, $E_{\rm{break}}$, in the model spectrum is mainly governed by the CASE 1 particles that are confined and escape from the cocoon (Figure \ref{f6}(b)). If the acceleration is sufficiently efficient, it would appear at the Hillas energy, { 
\begin{equation}
E_{\rm{break}}\approx E_H\approx0.9~{\rm{EeV}}\left(\frac{\langle B\rangle}{1\mu\rm{G}}\right)\left(\frac{0.5\mathcal{W}}{1{\rm kpc}}\right), 
\label{Ehillas}
\end{equation}
estimated with the mean magnetic field strength, $\langle B\rangle$, and the transverse size of the cocoon, $\mathcal{W}$. Note that $E_H$ is different from the {\it local} Hillas Energy, $E_{H,L_0}$, in Equation (\ref{EHL0}), which is defined by the local magnetic field strength, $B(x,y,z)$, and the coherent scale of turbulence, $L_0$.}

On the contrary, if the acceleration is slow, CRs may escape diffusively from the turbulent cocoon before they reach the Hillas energy, $E_H$; then the maximum energy is limited by escape via fast spatial diffusion. The sample trajectory shown in Figure \ref{f4} is for a CASE 1 particle that undergoes inefficient acceleration in the diffusive head region. The timescale for particles to diffuse across the cocoon is $t_{\rm{diff}}\approx(\mathcal{W}/2)^2/(c\lambda_f)$. The particles around $E_{\rm{break}}$ have $\lambda_f\gtrsim r_j$ and hence could cross the jet-backflow interface. Then, nGSA would be operative, and its acceleration timescale can be approximated as $t_{\rm{nGSA}}\approx\lambda_f/c\langle\Gamma\beta\rangle_{\rm acc}^2$ \citep[e.g.][]{kimura2018}, where $\beta\equiv u_z/c$ and $\langle\Gamma\beta\rangle_{\rm acc}$ is the volume-averaged value over the region of nGSA. Figure~\ref{f5}(a) shows that $t_D$ is smaller than $t_{\rm{nGSA}}$ for $E\gtrsim E_{\rm{break}}$ in Q44-r100, in which the jet flow is significantly decelerated. { Thus, in the case of slow acceleration, the maximum energy may be estimated from the condition $t_D\approx t_{\rm{nGSA}}$. For the fiducial model with $\lambda_f(p)\propto E$, this condition results in
\begin{equation}
E_{\rm{break}}\approx E_D\approx E_H \cdot \langle\Gamma\beta\rangle_{\rm{acc}}. \label{ED}
\end{equation}}

{The twelfth column of Table \ref{t1} lists the values of $E_H$, which are calculated using the volume-averaged magnetic field strength in the cocoon and the width of the cocoon at $z=l_{\rm{jet}}/2$ in simulated jets at $t_{\rm{end}}$.} The thirteenth column lists the values of $E_D$ at $t_{\rm{end}}$ using $\langle\Gamma\beta\rangle_{\rm{acc}}$, the volume-averaged value at the jet-spine in the upper half of the jet ($z=(1/2-1)l_{\rm{jet}}$) where the majority of CASE 1 particles are confined before they escape to the background (see Figure \ref{f4}). Both $E_H$ and $E_D$ approach to time-asymptotic values in our jet models, although their time evolution is not explicitly presented. In the FR-I models with large injection Lorentz factors, $\Gamma_j=11.1$ and $34.5$ ($\langle\Gamma\rangle_{\rm{spine}}\gtrsim6$), $E_{\rm{break}}\approx E_H$, indicating that RSA is efficient and the majority of particles reach $E_H$ before they escape. In contrast, in the models with $\Gamma_j=3.9$ ($\langle\Gamma\rangle_{\rm{spine}}\lesssim3$), {$\langle\Gamma\beta\rangle_{\rm{acc}}\sim 0.4 - 0.5$,} and hence $ E_{\rm{break}}\approx E_D\lesssim E_H$, implying that RSA may not be sufficiently efficient because the jet flow is only mildly relativistic. {Therefore, $E_{\rm{break}}\approx E_H$ may be adopted in our model spectrum for jets with $\langle\Gamma\rangle_{\rm{spine}}\gtrsim 6$, whereas for jets with $\langle\Gamma\rangle_{\rm{spine}}\lesssim 3$, $E_{\rm{break}}\approx E_D$ needs to be employed.}

The break energy, $E_{\rm{break}}$, is expected to depend on the properties of the jet and hence the jet model parameters such as $Q_j$, $r_j$, and the background density $\rho_b$, in our simulations. In Paper I, we found that the time-asymptotic value of $E_{\rm{break}}\approx E_H$ scales as $Q_j^{1/3}$ for FR-II jets.

{To elicit the approximate dependence of $E_H$ on $Q_j$ as well as $r_j$ and $\rho_b$, we here consider a relativistic jet injected into a {\it uniform} background medium, and assume an idealized cylindrical volume with radius $R_c$ and vertical length $l_c$ for the jet-induced cocoon. If a significant fraction of the kinetic energy of the jet dissipates into the internal energy of the cocoon, the gas pressure of the cocoon can be approximated as $ P_c \approx ( Q_j t)/(\pi R_c^2 l_c)$. We take $R_c\approx u_r t$ and $l_c\approx u_{\rm head} t$; the radial expansion speed is given as $u_r\approx (P_c/\rho_b)^{1/2}$ from the ram pressure balance, and the head advance speed is given as $u_{\rm head}\approx \sqrt{\eta_r}c$ with $u_j\approx c$ and $\eta_r\approx Q_j/(\pi r_j^2 c^3 \rho_b) \ll 1$ (see Section \ref{s2.1}). Using these relations, we obtain the scaling for the cocoon pressure:
\begin{equation}
P_c \propto \left(\frac{Q_j} {r_j^2}\right)^{3/4} \rho_b^{1/4} \left(\frac{t}{t_{\rm cross}}\right)^{-1}.
\end{equation}
Then, the width of the cocoon can be estimated as
\begin{equation}
\mathcal{W} \approx 2u_r t \propto \left(\frac{Q_j} {r_j^2}\right)^{-1/8} \rho_b^{1/8} \left(\frac{t}{t_{\rm cross}}\right)^{1/2} r_j. \label{w}
\end{equation}
Moreover, in our prescription for $B$, the magnetic field is stronger if the turbulence energy is larger and the gas pressure is larger (see Section \ref{s3.1}); hence, we may approximate the mean magnetic field in the cocoon as
\begin{equation}
\langle B\rangle\propto P_c^{1/2}\propto \left(\frac{Q_j} {r_j^2}\right)^{3/8}\rho_b^{1/8} \left(\frac{t}{t_{\rm cross}}\right)^{-1/2}.
\end{equation}
Then, $E_H$ scales as
\begin{equation}
E_H\propto\mathcal{W}\cdot\langle B\rangle\propto Q_j^{1/4}r_j^{1/2}\rho_b^{1/4}. \label{EH}
\end{equation}
We note that while the width increases in time as $\mathcal{W}\propto t^{1/2}$, the magnetic field strength decreases as $\langle B\rangle\propto t^{-1/2}$. As a consequence, $E_H$ is independent of time.} As a matter of fact, this is one of the reasons why the shape of the energy spectra in our simulations approaches a time-asymptotic form, as mentioned above.

In Table \ref{t1}, the Hillas energy, $E_H$, scales approximately as $Q_j^{1/4}$, rather than as $Q_j^{1/3}$. This could be because the FR-I jets considered here are more compact with smaller spatial dimensions than the FR-II jets of Paper I. On the other hand, the dependence on $r_j$ is not obvious, possibly because the jets propagate through the background of different stratifications (that is, different $r_{\rm core}$) in models with different $r_j$.

Observed RG jets are commonly quantified with radio or X-ray luminosities that may be related to $Q_j$, but other properties such as $r_j$ and $\rho_b$ are not easily estimated from observations. 
{Hence, we propose an expression for $E_{\rm{break}}$ for the model spectrum of UHECRs from FR-I jets as
\begin{equation}
E_{\rm{break}}\approx \varphi E_H \sim 45~{\rm EeV} \times \varphi \cdot \xi \cdot \left(\frac{Q_j}{Q_0}\right)^{\alpha}, \label{ebreak}
\end{equation}
where $Q_0=3.5\times 10^{44}~{\rm erg s^{-1}}$ and $\alpha=1/4$. The first fuzzy factor, $\varphi\sim\min[1,\langle\Gamma\beta\rangle_{\rm{acc}}]$, becomes $\varphi\approx 1$ for high-power jets with $\langle\Gamma\rangle_{\rm spine} \gtrsim 6$, while $\varphi\sim \langle\Gamma\beta\rangle_{\rm{acc}}\sim 0.4-1$ for low-power jets with $\langle\Gamma\rangle_{\rm spine} \lesssim 6$, based on the jet simulations considered here. This factor could be roughly estimated if the jet-spine structure is resolved in observations. The second fuzzy factor, $\xi\approx ({r_j}/{r_0})^{1/2}({\rho_b}/{\rho_0})^{1/4}$, where $r_0=100~{\rm pc}$, $\rho_0 = 3\times10^{-25}~{\rm g cm^{-3}}$. Although it would be difficult to constrain $\xi$ from observations of RG jets, $\xi\sim 0.4-1$ for the jet models considered here.} The above expression may be applied to FR-II jets as well, with $\alpha=1/3$ and $\varphi\sim1$.

\subsection{Model Dependence of Energy Spectrum}\label{s4.5}

\begin{figure}[t]
\vskip 0 cm
\hskip -0.3 cm
\includegraphics[width=1.03\linewidth]{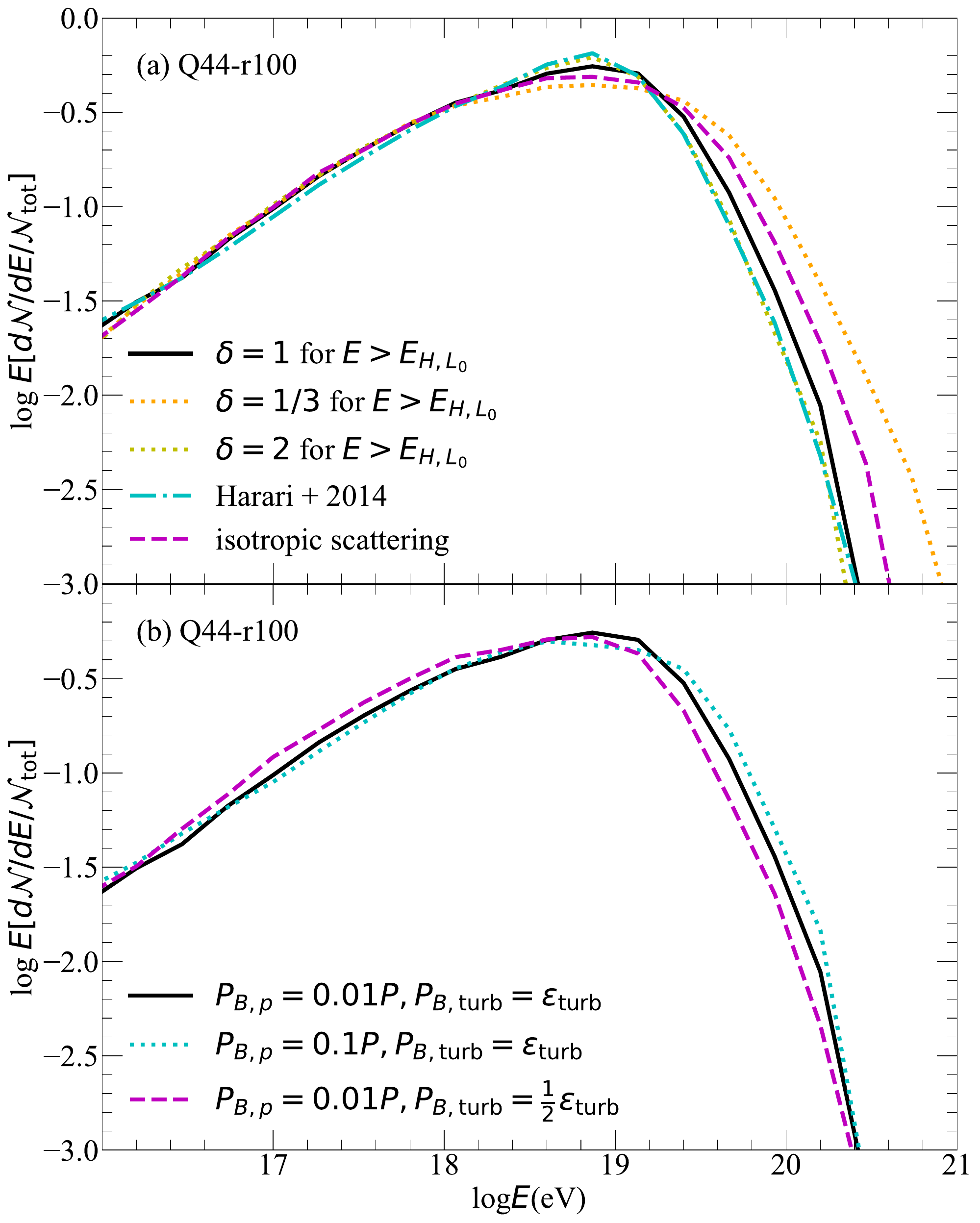}
\vskip -0.2 cm
\caption{Panel (a): Energy spectra with different modelings for the scattering of CR particles in the Q44-r100 model. The diffusive scattering model by \citet{harari2014} is given in Equation (\ref{harari}). The purple dashed line shows the case where fully isotropic scattering is adopted. Panel (b): Energy spectra with different modelings for the magnetic field distribution in the Q44-r100 model. Here, the spectrum of the fiducial case (black solid line) is the same as that in Figure \ref{f7}(b).}\label{f8}
\end{figure}

The spectrum in Equation~(\ref{powerexp}) is the result of MC simulations based on a number of modelings described in Section \ref{s3}. {Here, we examine the dependence on particle scattering and magnetic field models, as those could affect the resulting spectrum.}

In Figure~\ref{f8}(a), the energy spectra with $\delta=1/3$ (yellow dotted) and $2$ (light green dotted) for $E>E_{H,L_0}$ are compared to the fiducial spectrum with $\delta=1$ (black solid) for the Q44-r100 model. Below $E_{H,L_0}\sim1$~EeV, the three spectra are basically identical. Above $E_{H,L_0}$, the spectrum with $\delta=1/3$ extends to higher energies, while the spectrum with $\delta=2$ falls off at lower energies. This is because with $\delta=1/3$, high-energy particles suffer more scatterings compared to the Bohm model with $\delta=1$, and hence reach higher energies. In contrast, with $\delta=2$, it works in the opposite way. {The cyan dot-dashed line shows the spectrum for the diffusion model in Equation (\ref{harari}) proposed by \citet{harari2014}, which almost overlaps the spectrum with $\delta=2$, as expected.} In the same panel, the energy spectrum of the fully isotropic scattering model is also shown with a purple dashed line. With isotropic scatterings rather than restricted random walks, the chance for high-energy particles to cross the shear interface increases, and nGSA is enhanced. Hence, the spectrum extends to higher energies.

{Figure~\ref{f8}(b) shows how the magnetic field model affects the resulting spectrum. The black solid line draws the spectrum of the fiducial model with $\mathcal{E}_{B,{\rm turb}}=\mathcal{E}_{\rm turb}$ and $P_{B,P}=0.01P$; the cyan dotted line plots the spectrum with $\mathcal{E}_{B,{\rm turb}}=\mathcal{E}_{\rm turb}$ and $P_{B,P}=0.1P$, and the purple dashed line plots the spectrum with $\mathcal{E}_{B,{\rm turb}}=(1/2)\mathcal{E}_{\rm turb}$ and $P_{B,P}=0.01P$. This comparison demonstrates that the spectrum of escaping UHECRs would shift to higher energies if the magnetic field is stronger (cyan dotted), while it works in the opposite way if the magnetic field is weaker (purple dashed).}

Figure~\ref{f8} implies that the predicted energy spectrum generated in RG jets is, to a certain extent, contingent upon the modeling details of physical processes, including the characteristics of magnetic turbulence and particle scattering.

\section{Summary}\label{s5}
 
In paper I, we performed RHD simulations to reproduce the dynamics of relativistic jets in {\it FR-II} RGs and MC simulations to track the transport and energization of CR particles in the simulated jet-induced flows. Following the same numerical approach of RHD and MC simulations, we here investigate the jet flows and the acceleration of UHECRs in {\it FR-I} RGs.

In FR-I jets with smaller power $Q_j$ and smaller Lorentz factor $\Gamma_j$, the jet-spine gets decelerated more significantly. Compared to high-power FR-II jets, these low-power jets develop broader cocoons and more diffusive head regions, containing more numerous shocks and turbulence but weaker velocity shear. 

The main results related to particle acceleration are summarized as follows:

1. DSA at shocks, TSA by turbulence, and RSA (both GSA and nGSA) at shear operate for the acceleration of UHECRs. While DSA is dominant in energizing CRs up to $E\sim1$~EeV, RSA becomes increasingly important beyond $E\sim1$~EeV. TSA plays a secondary role.

2. The time-integrated, cumulative energy spectrum of all particles escaping from the jet system approaches a time-asymptotic form. It can be modeled by a double power law with an exponential cutoff as given in Equation~(\ref{powerexp}).

3. The break energy of power laws, $E_{\rm{break}}$, is either fixed by the Hillas confinement condition or limited by particle escape from the cocoon via fast spatial diffusion. It depends on jet parameters and may be modeled as Equation~(\ref{ebreak}).

4. The power-law slopes are on average $s_1\approx-0.6$ below $E_{\rm break}$ and $s_2\approx-2.6$ above $E_{\rm break}$, and show only a weak dependence on jet models. The hard spectrum, $E^{-0.6}$, for $E<E_{\rm{break}}$ is thought to result from multiple episodes of DSA and RSA in the jet system. In contrast, the steep spectrum, $E^{-2.6}$, for $E>E_{\rm{break}}$ is caused by the particle confinement and escape from the elongated cocoon.

5. The exponential cutoff energy is given as $\sim E_{\rm{break}}\langle\Gamma\rangle_{\rm{spine}}^2$, where $\langle\Gamma\rangle_{\rm{spine}}$ is the mean Lorentz factor of the jet-spine. It is governed by nGSA at the shear interface, which increases the CR energy by a factor of $\sim\langle\Gamma\rangle_{\rm{spine}}^2$.

The model spectrum derived for FR-I jets in this work, along with the one for FR-II jets in Paper I, may be utilized to investigate the contributions of nearby RGs as well as cosmological populations of RGs to UHECRs observed at Earth. This would help elucidate the origins and mechanisms behind the production of UHECRs. We leave such a study as a future work.

\begin{acknowledgments}
The authors thank the anonymous referee for constructive comments. This work was supported by the National Research Foundation (NRF) of Korea through grants 2020R1A2C2102800, 2023R1A2C1003131, and RS-2022-00197685. Some of simulations were performed using the high performance computing resources of the UNIST Supercomputing Center.
\end{acknowledgments}

\bibliography{RadioGalaxy}{}
\bibliographystyle{aasjournal}

\end{document}